\documentclass[3p,times]{elsarticle}

\usepackage{amssymb}
\usepackage{amsmath}
\usepackage{lineno}
\journal{Nuclear Instruments and Methods in Physics Research A}
\begin{document}

\begin{frontmatter}

%% Note: \pmbanner before the actual title
\title{ The measuring systems of the wire tension for the MEG II Drift Chamber by means of the resonant frequency technique }

\author[infnPI]{A. M.~Baldini}
\author[uniPI,infnPI]{H.~Benmansour}
\author[uniRM,infnRM]{G.~Cavoto}
\author[uniPI,infnPI]{F.~Cei}
\ead{fabrizio.cei@unipi.it}
\author[infnPI]{M.~Chiappini}
\author[infnPI]{G.~Chiarello}
\author[infnLE]{C.~Chiri}
\author[uniLE,infnLE]{G.~Cocciolo}
\author[infnLE]{A.~Corvaglia}
\author[uniRM,infnRM]{F.~Cuna}
\author[uniPI,infnPI]{M.~Francesconi}
\author[infnPI]{L.~Galli}
\author[infnLE]{F.~Grancagnolo}
\author[infnPI]{M.~Grassi}
\author[uniRM,infnRM]{M.~Meucci}
\author[infnLE]{A.~Miccoli}
\author[uniPI,infnPI]{D.~Nicol\'{o}}
\author[uniLE,infnLE]{M.~Panareo}
\author[uniPI,infnPI,PSI]{A.~Papa}
%\author[uniLE,infnLE]{A.~Pepino}
\author[uniLE,infnLE]{C.~Pinto}
\author[infnPI]{F.~Raffaelli}
\author[infnRM]{F.~Renga}
%\author[uniRM,infnRM]{E.~Ripiccini}
\author[infnPI]{G.~Signorelli}
\author[uniBA]{G.F.~Tassielli}
\author[uniPI,infnPI]{A.~Venturini}
\author[uniPI,infnPI]{B.~Vitali}
\author[infnRM]{C.~Voena}
\cortext[cor]{Corresponding author}
\address[uniRM]{Dipartimento di Fisica dell'Universit\`{a} \lq\lq La Sapienza\rq\rq,  Piazzale A. Moro, 00185 Roma, Italy}
\address[infnRM]{INFN Sezione di Roma, Piazzale A. Moro, 00185 Roma, Italy}
\address[uniLE]{Dipartimento di Matematica e Fisica "Ennio De Giorgi" - Universit\`{a} del Salento, Via Arnesano, Lecce, Italy}
\address[infnLE]{INFN Sezione di Lecce, Via Arnesano, Lecce, Italy}
\address[infnPI]{INFN Sezione di Pisa, Largo B. Pontecorvo 3, 56127, Pisa, Italy}
\address[uniPI]{Dipartimento di Fisica dellUniversit\`{a} di Pisa, Largo B. Pontecorvo 3, 56127 Pisa, Italy}
\address[uniBA]{Dipartimento di Fisica dellUniversit\`{a} di Bari, Campus Universitario, Via Amendola 173, 70125 Bari, Italy}
\address[PSI]{Paul Scherrer Institut PSI, 5232 Villigen, Switzerland}

\begin{abstract}
The ultra-low mass cylindrical drift chamber designed for the MEG II experiment is a challenging apparatus made of 1728 $\phi=20~\mu{\rm m}$ gold plated tungsten sense wires, 7680 $\phi=40~\mu{\rm m}$ and 2496 $\phi=50~\mu{\rm m}$ silver plated aluminium field wires. 
Because of electrostatic stability requirements all the wires have to be stretched at mechanical tensions of $\sim25$, $\sim19$ and $\sim29~{\rm g}$ respectively which must be controlled at a level better than $0.5~{\rm g}$. 
This chamber is presently in acquisition, but during its construction $\sim100$ field wires broke, because of chemical corrosion induced by the atmospheric humidity. 

On the basis of the experience gained with this chamber we decided to build a new one, equipped with a different type of wires less sensitive to corrosion. The choice of the new wire required a deep inspection of its characteristics and one of the main tools for doing this is a system for measuring the wire tension by means of the resonant frequency technique, which is described in this paper. 
The system forces the wires 
to oscillate by applying a sinusoidal signal at a known frequency, and then measures the variation of the capacitance between a wire and a common ground plane as a function of the external signal frequency. We present the details of the measuring system and the results obtained by scanning the mechanical tensions of two samples of MEG II cylindrical drift chamber wires and discuss the possible improvements of the experimental apparatus and of the measuring technique.
\end{abstract}

\end{frontmatter}

%\linenumbers

\section{Introduction}
The Cylindrical Drift Chamber (from now on: CDCH)~\cite{Chiappini,ChiarelloCD,SingleHit}
is the most innovative part of the MEG II experiment~\cite{Meg2Detector}, an upgraded version of the MEG experiment~\cite{MegDetector} at Paul Scherrer Institute~\cite{PSI} of Villigen (Switzerland), which established the world most stringent upper limit on the $\mu^{+} \rightarrow {\rm e}^{+} \gamma$~\cite{MegFinalPaper} and a significant upper limit on the $\mu \rightarrow {\rm e}^{+} \gamma \gamma$ decay too~\cite{Megmue2gamma}. 
This new chamber is a unique volume detector, of $1.93~{\rm m}$ length and $34~{\rm cm}$ inner diameter, equipped with $1728$ $\phi=20~\mu{\rm m}$\footnote{$\phi$ is the international symbol standing for \lq\lq diameter\rq\rq.} gold plated tungsten sense wires, $7680$ $\phi=40~\mu{\rm m}$ and $2496$ $\phi=50~\mu{\rm m}$ silver plated aluminium wires, used to define the electric field configuration within the gaseous volume and at its boundaries. This huge number of wires forms a dense matrix of approximately squared drift cells, whose size ranges from $\sim5.8~{\rm mm}$ at the chamber centre to $\sim8.7~{\rm mm}$ close to the chamber ends. The detector is immersed in a longitudinally varying solenoidal magnetic field produced by a superconducting magnet (COBRA, COnstant Bending RAdius), arranged to rapidly sweep away from the chamber region the positrons emitted with small longitudinal momenta; the maximum value of the magnetic field is $1.26~{\rm T}$ at the detector centre and is reduced by about a factor two at the longitudinal ends of the chamber. 
The chamber and the other subdetectors of the MEG II experiment are equipped with an innovative integrated trigger and acquisition system and with a custom made high performance readout electronics, extensively described in~\cite{MEGIIDAQ,MEGIITrigger}.

A positron emitted in a $\mu^{+} \rightarrow {\rm e}^{+} \gamma$ decay, whose energy is $52.83~{\rm MeV}$, crosses on average an amount of material equivalent to $1.6 \times 10^{-3}$ radiation lengths $X_{0}$; the expected momentum and angular resolutions at that energy are $\sim100~{\rm keV}$ and $\sim6~{\rm mrad}$ respectively. 

The use of very thin wires is mandatory to reduce as much as possible the energy resolution degradation due to energy loss and multiple scattering. On the other hand, particular care must be put during the handling and assembly operations of such wires: the wiring operation is automatically performed by a computer-controlled wiring robot~\cite{ChiarelloWR} which solders the wires on metallic pads deposited on printed circuit boards (from now on: PCBs) and builds several multi-wire layers; then the layers are extracted, placed in storage and transport frames, moved to the assembly site and finally installed on the chamber end plate supports. The silver coating on aluminium wires was used to protect the wire aluminium core and to make the soldering easier and more stable.   

Since the potential difference between anode and cathode wires in the chamber is $\sim1.5~{\rm kV}$, wires with charges of opposite sign are subject to a strong electrostatic attraction, which can cause geometrical modifications and short circuits. An appropriate mechanical tension is therefore needed to counterbalance the electrostatic force and prevent the wires to collapse. 

A very critical point is the possible chemical corrosion of the wires, due to the exposure to room humidity, which can induce wire breakings when the wires are stretched. During the chamber construction this problem arised and caused the breakage of $\sim100$ cathode wires, almost all with $\phi=40~\mu{\rm m}$. Because of the huge number of cathode wires ($\sim8000$), the effect of few tens of broken cathode wires on the CDCH performances is absolutely negligible; however, the presence of wire fragments within the chamber is dangerous because of the strong electric fields inside it. Therefore all the broken wires were carefully removed, the chamber construction was completed and it is now in acquisition at PSI. However, in case of further breakages we would be forced to stop the acquisition, open the chamber in a controlled atmosphere, remove all fragments and close and seal the chamber again. This operation is delicate and requires several weeks. 
We studied the corrosion problem in deep details, both experimentally and phenomenologically, as discussed in a previous paper~\cite{BrokenWires} and developed a strategy to minimise the risk of new breakages. Our detailed analysis shows that the chemical corrosion of the wires was originated by the formation of cracks in the silver coating, mainly during the last phase of the wire production, the so-called \lq\lq ultrafinishing\rq\rq~procedure, and that the breakage probability is much smaller for the thicker $\phi=50~\mu{\rm m}$ wires. Therefore, we decided to build a new chamber (while the old one is in acquisition) equipped with a different type of cathode wires, less sensitive to chemical corrosion. 
Several possibilities were explored and our present preference, not yet a final choice, is in favour of $\phi=50~\mu{\rm m}$ silver plated aluminium wires, without ultrafinishing procedure, which is unavoidable for a $\phi=40~\mu{\rm m}$, but not strictly needed for a $\phi=50~\mu{\rm m}$ wire. The motivations for this choice and the constraints which must be satisfied by the selected wire are discussed in~\cite{BrokenWires}, one of the most important being the long-term mechanical stability of the wire when it is stretched to compensate the electrostatic attractant force. The wire length and tension must remain stable on time scales of at least several months, avoiding significant deformations of the wire and maintaning a good electrical contact with the electronic boards.      
 
In this paper we show a technique for measuring the mechanical tension of the wires based on the relationship between the tension itself and the resonant frequency of an elastic string. We describe the experimental apparatus, the acquisition and analysis software and the results of the measurements on two samples of wires made by different aluminium alloys. We discuss the evolution of the wire mechanical tension of these two samples, the offline corrections to be applied and the possibility of using them in the new CDCH. We discuss also how the performances of the measuring device can be improved by using a new electronics, less sensitive to room temperature, or by moving the apparatus in a controlled environment.    

\section{Characteristics of the wires used in the MEG II drift chamber}
In the MEG II CDCH, three kinds of wires are used:
\begin{itemize}
    \item the anode (sense) wires, which collect the ionisation electrons generated by the charged particles traversing the drift chamber gas;
    \item the cathode (field) wires, which define the drift cells of the chamber;
    \item the guard wires, which define the electric field configuration near the edges of the active  volume.
\end{itemize}
Several wire types, manufactured by different companies (Alloy Wire International~\cite{alw}, California Fine Wire Company  (from now on: CFW)~\cite{cfw} and  Luma Metall Fine Wire Products~\cite{lmfw}) were considered, taking into account multiple requirements, as density, radiation length $X_0$, equivalent radiation length of all wires and of wires and CDCH filling gas, and multiple scattering angular deflection $\Theta_{MSC}$ for a $52.83~{\rm MeV}$ positron track. 
The best performance for field and guard wires was obtained with the aluminium ${\rm Al}5056$ alloy from the CFW company, composed by aluminium (94.6\%), magnesium (5.01\%), iron (0.14\%), silicon (0.1\%) and traces of other elements. The wires  are produced in different 4-inch spools by using an initial wire drawing procedure which decreases the wire diameter to $55~\mu{\rm m}$ in different phases; then an ultra-finishing procedure is applied to reach the design diameter of $40~\mu{\rm m}$ or $50~\mu{\rm m}$. The silver coating is performed by electrochemical deposition before the ultra-finishing stage; therefore this coating is exposed to a mechanical stress which can produce defects on the wire surface. 
 
\begin{figure}[htb]
	\centering
	\includegraphics[width=1\linewidth]{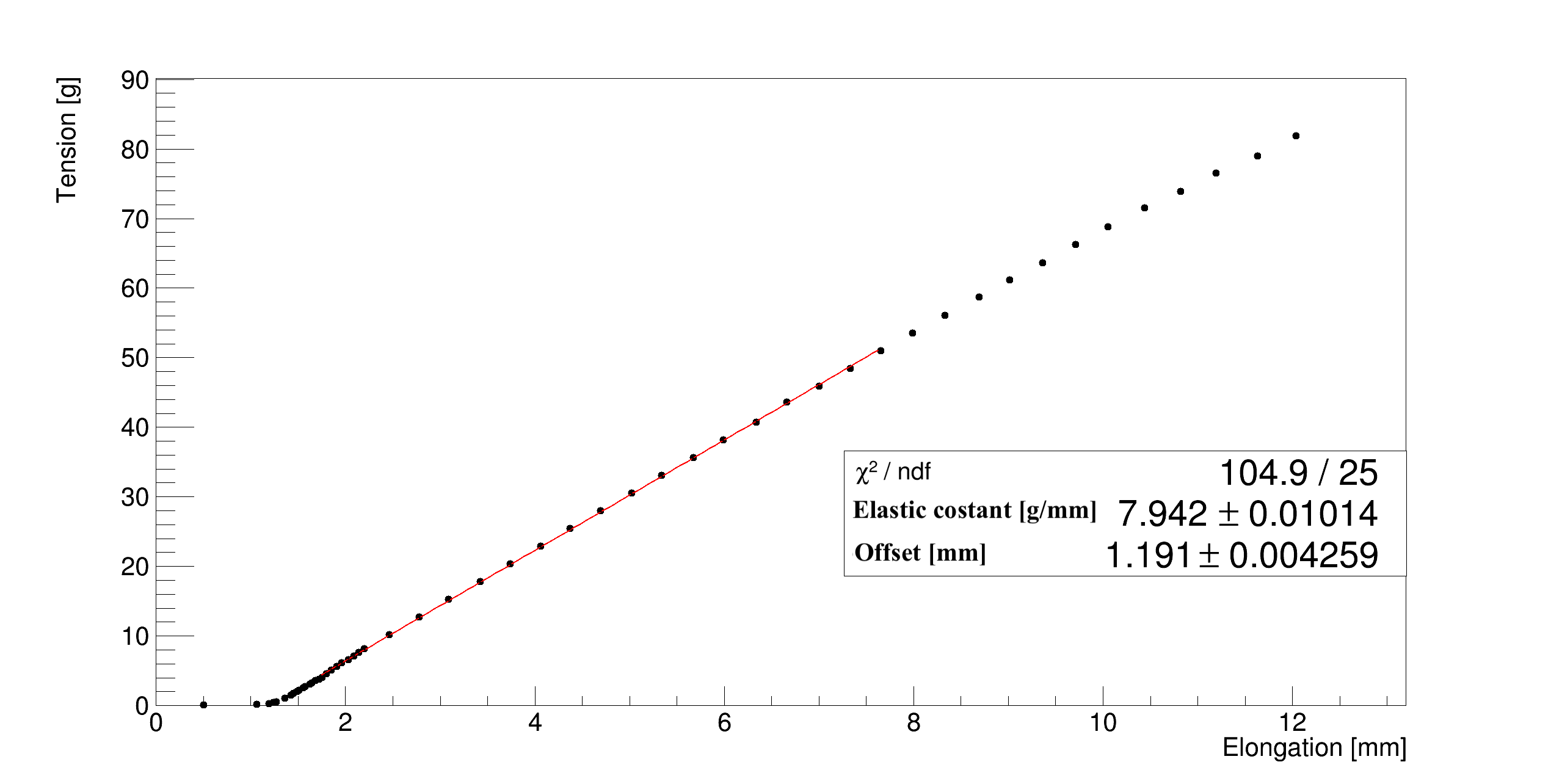}
	\caption{A stress-strain curve for a Al5056-silver plated $\phi=50~\mu{\rm m}$ wire.}
	\label{fig:stres50}
\end{figure}

The density and the resistivity of the ${\rm Al}5056$ alloy are $\delta=2.64~{\rm g/cm^3}$ and $\rho \simeq 20~\Omega \times {\rm m}$ respectively; the length of the wires is $\sim1.93~{\rm m}$. An example of a stress-strain\footnote{\lq\lq Stress\rq\rq~indicates the longitudinal tension force acting on a mechanical wire, sometimes divided by its the cross section; \lq\lq Strain\rq\rq~indicates the elongation of the wire when it is exposed to the tension force.} curve for a $\phi=50~\mu{\rm m}$ aluminium wire is shown in figure~\ref{fig:stres50}.
Pure gold plated tungsten is used for anode wires, whose diameter is $20~\mu{\rm m}$; the anode wire length is $\sim1.93~{\rm m}$ too. The anode wire material has a density 
$\delta=19.27~{\rm g/cm^3}$ and a resistivity $\rho=175~\Omega \times {\rm m}$. This type of wire allows us to use a readout electronics with lower gain and bandwidth (and hence lower noise) compared to other wires considered in order to be able to use the cluster counting technique~\cite{ClusterCount}. Differently from the aluminium wires, the tungsten wires have a long plastic phase, so that their breaking length is much larger than the elastic limit; in any case the elastic limit of these wires corresponds to $\sim10.2~{\rm mm}$, almost twice the working elongation, as explained at the end of the paragraph.   

The mechanical parameters of the silver plated Al$5056$ and gold plated tungsten wires selected for the CDCH are summarised in table~\ref{tab:wireparam}. 
\begin{table*}[htb]
	\centering
	\caption{Mechanical parameters of the MEG II CDCH wires.}
	\begin{tabular}{|l|cc|cc|c|}
		\hline	
		&\multicolumn{2}{|c|}{Elastic limit} &\multicolumn{2}{|c|}{Breaking point}&Elastic Modulus \\
                   \hline	
		Wire type & Elongation & Tension & Elongation & Tension& $\left( {\rm g/mm/m} \right)$\\
		&$\left( {\rm mm/m} \right)$ & $\left( {\rm g} \right)$ & $\left( {\rm mm/m} \right)$ &$ \left( {\rm g} \right)$& \\
		\hline
		\hline
		20~$\mu$m W & $5.30\pm 0.05$ & $62.0\pm 0.7$ & $19.7\pm 0.3$ & $86.0\pm 2.5$ & $11.92\pm 0.09$\\
		40~$\mu$m Al & $4.81\pm 0.14$ & $44.3\pm 0.8$ & $5.82\pm 0.33$ & $50.0\pm 2.5$ & $9.20\pm 0.13$\\
		50~$\mu$m Al & $4.90\pm 0.04$ & $69.5\pm 0.5$ & $5.58\pm 0.31$ & $75.3\pm 3.9$ & $14.17\pm 0.20$\\
		\hline
	\end{tabular}
	\label{tab:wireparam}	
\end{table*}
The operating point was chosen at an absolute elongation of $5.2~{\rm mm}$ with respect 
to the rest length, corresponding to about $60\%$ of the elastic limit of all wires. This value was obtained by means of detailed Garfield~\cite{Garfield} based simulations of the electrostatic stability of the whole chamber and confirmed experimentally with the chamber in operation. 
\section{Wire tension measurement theory}
The mechanical tension of the wires is one of the fundamental parameters that determines the correct operation of a drift chamber since it avoids wire deformations and short circuits between anode and cathode wires due to their potential difference. Performing a measurement of the mechanical tension of the wires before and during the chamber construction is crucial in order to understand the operation capabilities of the drift chamber and to figure out possible weak points in its construction. The wire mechanical tension needs to be monitored also after the installation on the chamber and when the wires have been stretched to the elongation needed to ensure the electrostatic stability and kept stable at a level $< 0.5~{\rm g}$. A precise measurement of the chamber dimensions is performed periodically at PSI by means of an optical survey. 

The indirect measurement of the wire mechanical tension is based on the method of the resonant oscillation induced on the wires~\cite{KLOE2}, already used by other high energy physics experiments~\cite{KLOE1,LHCB}. The  nominal resonant frequency $\nu_r$ of a wire is given by:
\begin{equation} 
\nu_r=\frac{1}{2L}\sqrt{\frac{T}{\mu}}
\label{eq:fwt}
\end{equation}
where $L$ is the wire length $\left( \sim 1.93~{\rm m} \right)$, $T$ is the wire tension and  $\mu$ is the wire linear mass density.

The nominal resonance frequencies for the three different wires used in the MEG II drift chamber and their working tensions are reported in table~\ref{tab:reson}; all wires have a nominal resonance frequency $\sim\left( 50\textendash55 \right)~{\rm Hz}$\footnote{Different aluminium alloys have different densities and then different resonance frequencies at fixed mechanical tension. For instance Al$2024$ alloy has a density $\delta=2.78~{\rm g/cm^3}$; the corresponding resonant frequency for $50~\mu$m Al (Ag) wires is $56.7~{\rm Hz}$ for a mechanical tension of $29.6~{\rm g}$.} 
\begin{table}[htb]
	\caption{Nominal resonance frequencies for the three types of wires. The linear density values take into account the composition of Al$5056$ and the thickness of wire coating.} 	
	\centering
	\begin{tabular}{|c|c|c|c|}
		\hline
		Wire type & $\mu~\left( 10^{-3}~{\rm g/m} \right)$ & $T~\left( {\rm g} \right)$ & $\nu_r~\left({\rm Hz} \right)$\\
		\hline
		\hline
		$20~\mu$m W (Au) &  6.04  & 24.1 & 51.7\\
		$40~\mu$m Al (Ag) &  3.81  & 19.3 & 57.8\\
		$50~\mu$m Al (Ag) &  5.79  & 29.6 & 58.0\\
		\hline
	\end{tabular}
	\label{tab:reson}
\end{table}
The technique for measuring the effective resonant frequency of a wire is based on the effects induced by an external harmonic electric field on the capacitance $C_{wg}$ of a system formed by a wire and a ground plane. The capacitance $C_{wg}$ is given by the equation:
\begin{equation} 
C_{wg}=\frac{2 \pi \varepsilon L}{\ln \left(\frac{4H}{d} - 1\right) }
\label{eq:cwt}
\end{equation}
where $H$ is the distance between the wire and the ground plane, $d$ is the wire diameter, $L$ is the wire length and $\varepsilon$ is the dielectric constant of the material in-between the wire and the ground plane. 

The ground plane is a metallic slab kept at a fixed position below the wire, while the wire is suspended on its ends and is free to oscillate as a violin chord. When a sinusoidal signal is applied to the wire, the wire is forced to oscillate and the distance between the wire and the ground plane changes with time, depending on the frequency of the external signal. Then, also the capacitance $C_{wg}$ varies with time and can be measured by using an auto oscillator circuit. When the distance between the wire and the ground plane increases, the capacitance becomes smaller (formula \ref{eq:cwt}) with respect to the situation when the wire is at rest, while when the distance decreases, the capacitance becomes larger. Then, the asymmetry between the capacitance measured when the wire is at the maximum and at the minimum distance from the ground plane is sensitive to the frequency of the external signal and is expected to reach its maximum value when this frequency matches the resonant frequency of the wire. 
\section{The measurement system}
\begin{figure}[htb]
	\centering
	\includegraphics[width=1.1\linewidth]{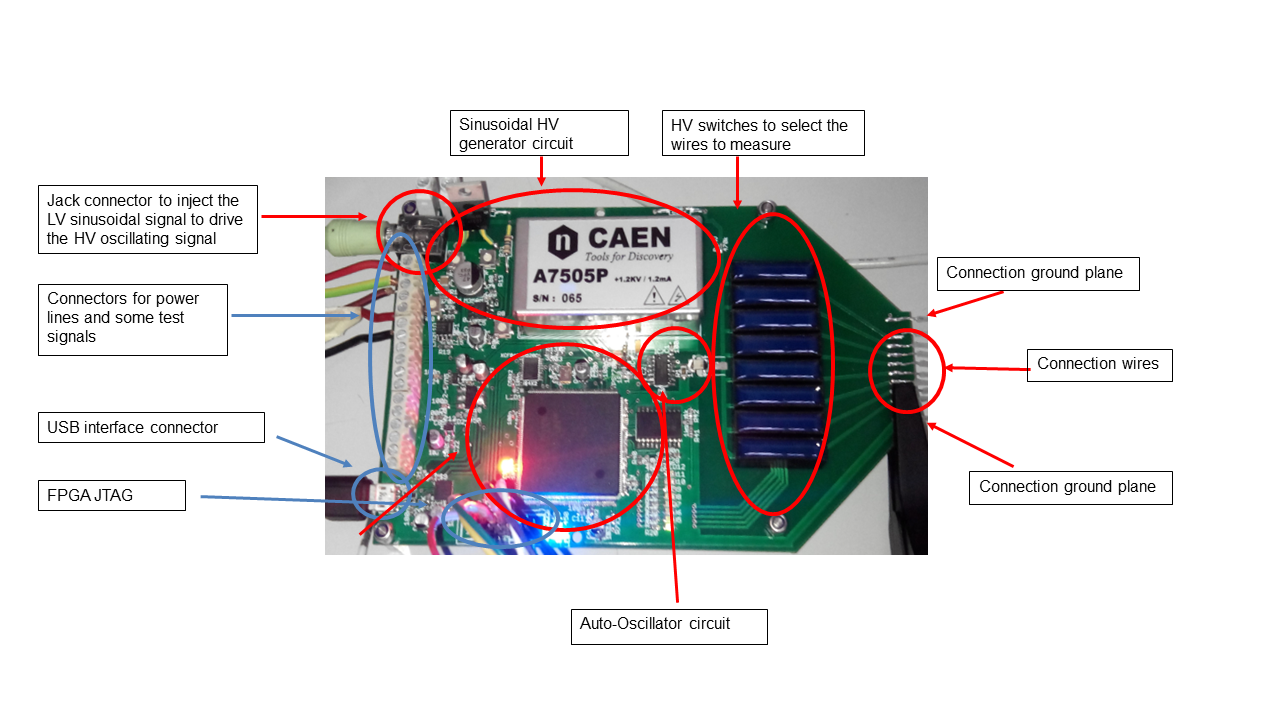}
         	\caption{The measurement system. The connections to the ground plane and to the wires are shown on the right.}
	\label{fig:meassyst}
\end{figure}

The measurement system is shown in figure \ref{fig:meassyst} and is formed by the following elements:
\begin{itemize}
	\item A sinusoidal High Voltage (from now on: HV) generator circuit with a CAEN A7505P chip;
         \item A sinusoidal Low Voltage (from now on: LV) generator (not shown in the figure), interfaced by means of a jack connector with the HV generator;
	\item a Ring type auto oscillator circuit for the measurement of the frequency variations;
	\item a Xilinx Spartan-3E FPGA to process the signals of the auto oscillator and of the LV generator and to handle the data;
	\item a group of HV switches to select the wires to measure;
	\item a USB port to connect the system with the DAQ computer through a serial interface; 
	\item a group of connectors for power lines and test signals.
\end{itemize}

\begin{figure}[htb]
	\centering
	\includegraphics[width=1\linewidth]{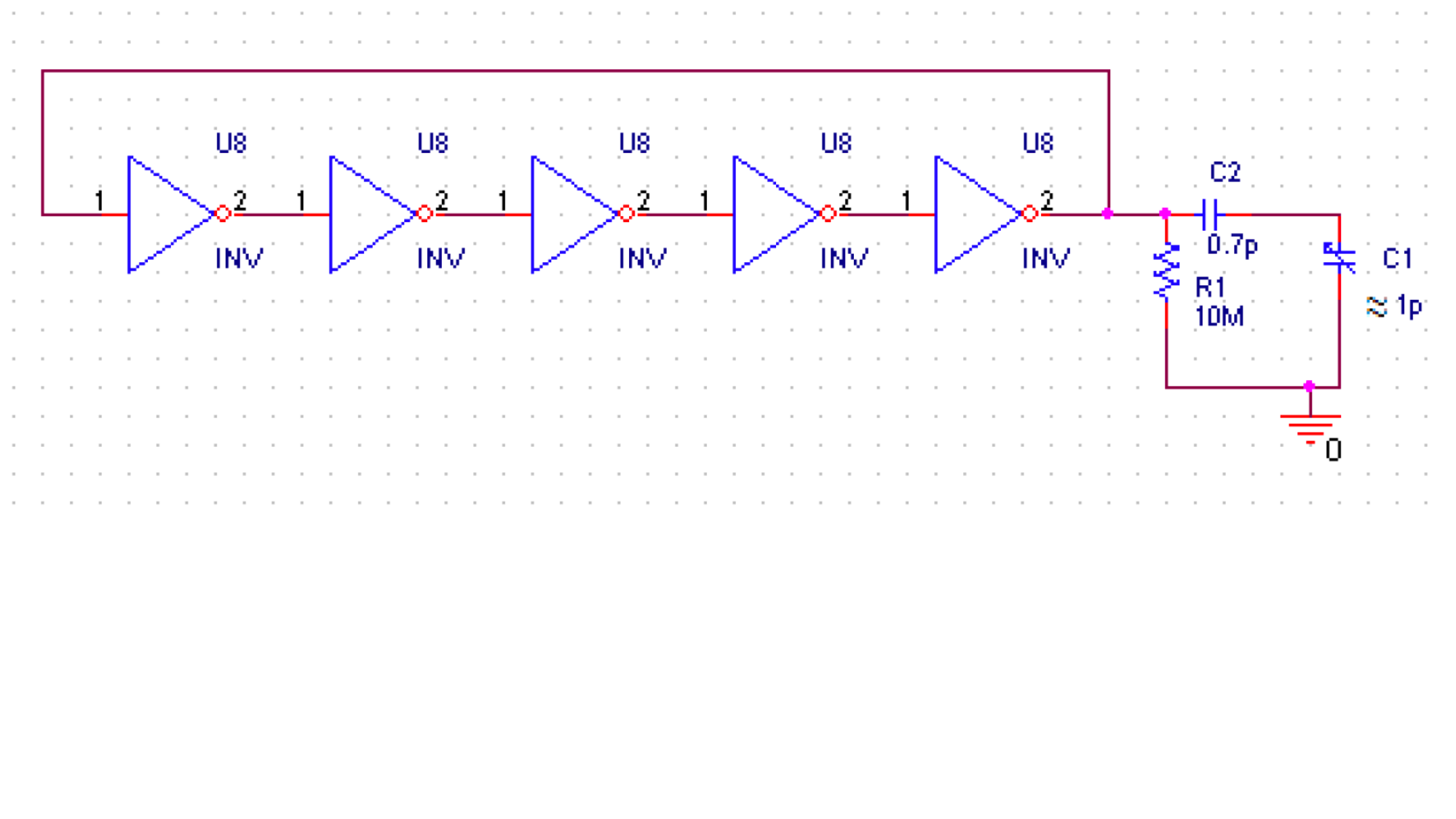}
	\vspace{-2.5cm}\caption{Electric scheme of the inverter chain of the Ring oscillator and equivalent circuit of the combination between the Ring oscillator, the wire under measurement and the ground plane. C$2$ and R$1$ are the intrinsic capacitance and the  resistance of the Ring oscillator, while C$1$ is the variable capacitance $C_{wg}$ between the wire under measurement and the ground plane. The effective value of $C_{wg}$ depends on the real distance between the wire and the ground plane, so that the value ${\rm C}_{1} \sim1~{\rm pF}$ is simply an indication of its expected order-of-magnitude.}
	\label{fig:eqcirc}
\end{figure}

The sinusoidal LV generator, controlled by the DAQ computer through an Ethernet connection, produces a $\left( 0\textendash4 \right)$ V amplitude and $\sim50~{\rm Hz}$ frequency signal, which is used to modulate a high frequency ($\sim40~{\rm MHz}$) signal, generated by means of an inverter chain. The LV signal is amplified from a few V to 
$\sim1~{\rm kV}$ in the HV generator and sent to the wire; the wire is then excited by a signal at the desired frequency of some tens of Hz and starts to oscillate, modifying the distance between the wire itself and the ground plane. The Ring type oscillator is connected to the wire under measurement; the combination between the oscillator, the wire and the ground plane is equivalent to an RC circuit, with two capacitances in series, that intrinsic of the oscillator and $C_{wg}$, as shown in figure~\ref{fig:eqcirc}.

The low frequency signal is also sent, before to be amplified, to a pair of comparators, interfaced with the FPGA chip, to form two square wave pulses and two synchronous gates. One of the square waves is inverted and shifted by $90^{\circ}$ in order to enhance the frequency asymmetry (see later) when one crosses the resonance condition. Since the two waves are inverted, one of the two gates is opened when the leading edge of the oscillating signal exceeds the discrimination threshold and is closed when the trailing edge goes under this threshold, while for the other gate the opposite situation takes place; then, one gate (from now on: gate $A$) is opened when the external pulse makes the distance between the wire and the ground plane to increase, and the other (from now on: gate $B$) when the external pulse makes this distance to decrease. 

Since a variation of the distance between the wire and the ground plane corresponds to a variation of their capacitance $C_{wg}$, when gate $A$ is open $C_{wg}$ is at its minimum value and when gate $B$ is open $C_{wg}$ is at its maximum value. Referring to the equivalent circuit in figure~\ref{fig:eqcirc} one can deduce that the Ring oscillator frequency is different during the opening times of gates $A$ and $B$. Note that the distance between the wire and the ground plane is a critical parameter of the measurement and must be carefully chosen: a too large distance would make the capacitance variations too small for a reliable measurement, while in case of a too small distance a large oscillation could cause a contact between the wire and the ground plane. We used an average distance of $\sim1.5$ cm; at this distance the capacitance $C_{wg}$ when the wire is at the rest position is $\sim18~{\rm pF}$, when the wire and the ground plane are at the maximum distance is $\sim16~{\rm pF}$ and when they are at the minimum distance is $\sim30~{\rm pF}$, so that the expected variation of the capacitance is $< 14~{\rm pF}$. 

The two gates have the same duration: $T_{A} = T_{B} = T = 1/\nu$, where $\nu$ is the frequency of the slow modulation signal ($\sim50~{\rm Hz}$), so that the number of Ring oscillations $N_{A}$ and $N_{B}$ measured during gates $A$ and $B$ is proportional to the frequencies $f_{A}$ and $f_{B}$ of the Ring oscillator at the opening times of the two gates. The frequency aysmmetry $\Delta f = f_{A} - f_{B}$ is therefore a measurement of the wire oscillation amplitude as a function of $\nu$. Because of the $90^{\circ}$ phase shift, the maximum $\Delta f$ occurs when $\nu = \nu_{r}$, i.e. in resonance conditions. The resonant frequency of the wire is therefore identified by searching for the maximum value of $\Delta f \left( \nu \right) = f_{A} \left( \nu \right) - f_{B} \left( \nu \right)$ as a function of $\nu$ by a frequency scan in a suitable range. Each $\Delta f$ measurement is repeated over several periods of the LV oscillating signal and the number of repetitions is chosen according to the required precision of the measurement. It is interesting to note that the typical value is $\Delta f \sim1~{\rm kHz}$, corresponding to a $\sim10^{-5}$ relative effect on a $\sim40~{\rm MHz}$ signal.

\begin{figure}[htb]
	\centering
	\includegraphics[width=1\linewidth]{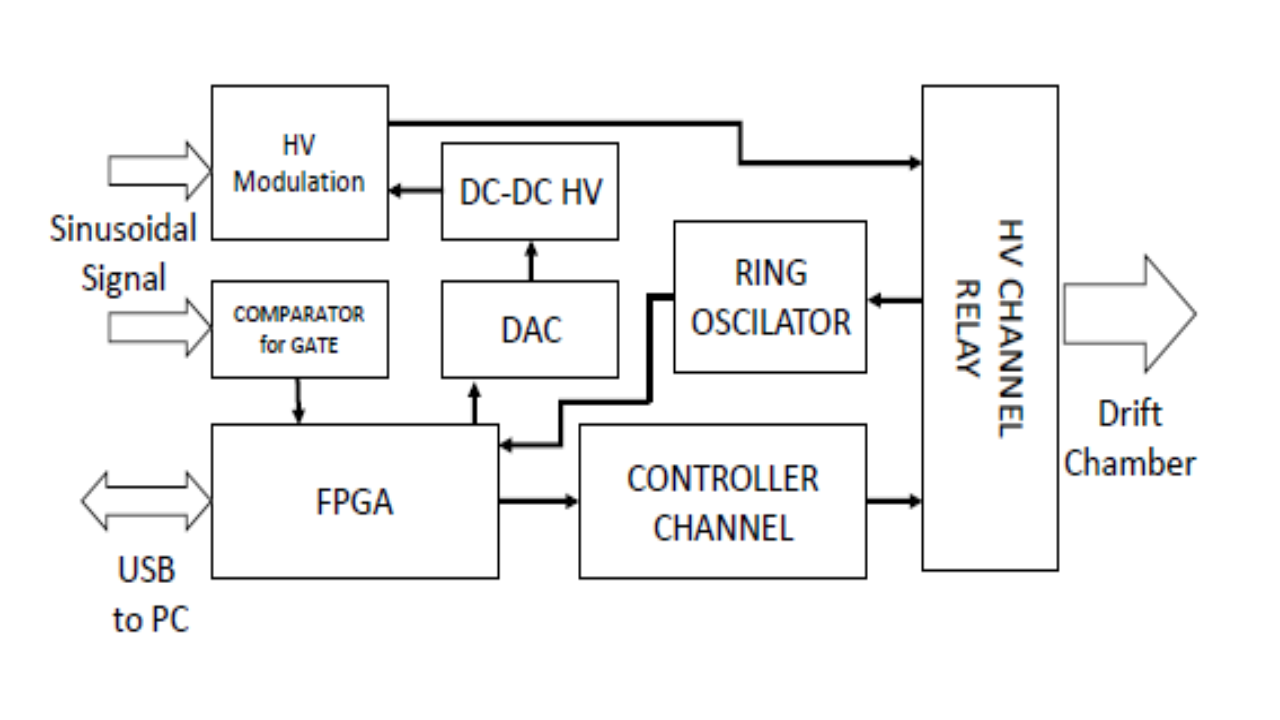}
	\caption{Block diagram of the measuring system}
	\label{fig:blocks}	
\end{figure}

Figure~\ref{fig:blocks} shows a block diagram of the measurement system. 
\section{DAQ and analysis software}
The DAQ code, originally written in Labview and then rewritten in C++ under ROOT environment to ensure higher flexibility, drives the low frequency generator and all interfaces between the measuring system and the acquisition computer. 

The DAQ software sends the excitation pulses to the wires in an adjustable frequency range, which can be tuned according to the expected region-of-interest for the resonance frequency. 
Since the resonant frequency is expected between $50~{\rm Hz}$ and $55~{\rm Hz}$, the excitation frequency scan is performed in the range $\left( 40\textendash60 \right)~{\rm Hz}$, at $0.25~{\rm Hz}$ steps. Fifty excitation pulses are sent for each measurement point and $<2$ minutes are needed to perform all measurements on a single wire. All data are automatically written and saved on disks, together with room information on temperature and humidity, needed to perform offline corrections. The analysis chain is largely automated and based on C codes working under ROOT environment; the data acquisition and the full analysis on a representative sample of six wires require less than $15$ minutes and can be easily performed on a day-by-day basis also on much larger samples. 

\begin{figure}[htb]
	\centering	
	\includegraphics[width=1\linewidth]{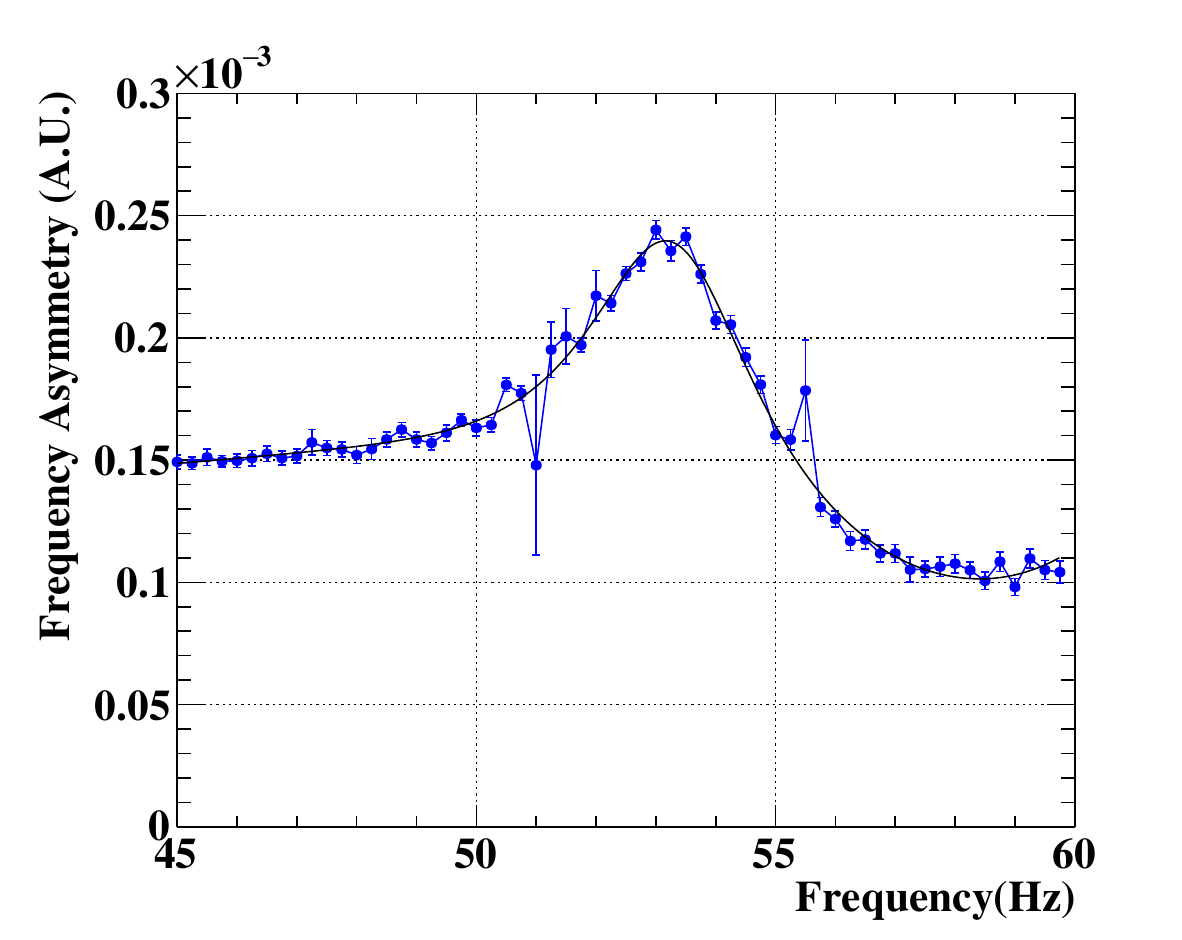}
	\caption{Example of the excitation curve of a MEG II CDCH wire. The Breit-Wigner behaviour close to the resonance at about $53~{\rm Hz}$ is clearly visible.}
	\label{fig:reso}	
\end{figure}

For each wire, the Ring oscillator frequency asymmetries\footnote{We remind that the frequency asymmetry of the Ring oscillator is a measurement of the amplitude of wire oscillations induced by the excitation signal.} of the fifty pulses corresponding to a given excitation frequency $\nu$ are averaged and their standard deviation is computed; then, the pairs $\left( \nu, \langle \Delta f \left( \nu \right) \rangle \right)$, where $\langle \Delta f \left( \nu \right) \rangle$ is the average frequency aysmmetry, are fit with a Breit-Wigner type curve plus a polynomial background:
\begin{equation}
\langle \Delta f \left( \nu \right) \rangle = \frac{A_{r}}{\left( 1 - \left( \frac{\nu}{\nu_{r}} \right)^{2}\right)^{2} + \left( \frac{\nu}{\Gamma_{r}} \right)^{2}} + \sum_{i=1}^{6} c_{i} \left( \nu - \bar{\nu} \right)^{i}
 	\label{eq:breit}
\end{equation}
In this formula $A_{r}$ is proportional to the frequency asymmetry at the resonance, $\nu_{r}$ is the resonant frequency, $\Gamma_{r}$ is the resonance width, $\bar{\nu}$ is fixed at the centre of the scan frequency range ($50~{\rm Hz}$) and $c_{i}$ are phenomenological coefficients. An example of a fit to an excitation curve is shown in figure~\ref{fig:reso}.
The mechanical tension of the wire is computed from the resonance frequency by inverting equation \ref{eq:fwt}.  
\section{Measurements and results}
The measuring system was in operation for more than one year in an INFN-Pisa laboratory. Despite the laboratory hadn't any temperature or humidity control infrastructure, the system operated in good conditions, with limited human interventions; however, as we will show later, the room conditions were monitored and a calibration of the temperature dependence of the resonant frequency was needed to obtain a reliable relationship between this frequency and the operation time of the instrument.

Here we show the results of two series of measurements which provided important information for the choice of the wire type for the new chamber. The wires were lodged in an 
aluminium frame and mechanically aligned with a $< 0.3^{\circ}$ relative precision. In the former series we tested the short term mechanical behaviour of a group of pure aluminium wires when they were glued on the PCBs instead of soldered and in the latter we measured the creeping and the long term stability of an other group of wires, made of a different aluminium alloy, fixed on the PCBs by means of the usual soldering.

\subsection{Measurements on glued wires}
As discussed in the introduction section, the formation of cracks in the silver coating facilitates the onset of the chemical corrosion. Therefore, one of the possible options for the wire in the new chamber is a $\phi=50~\mu{\rm m}$ aluminium wire without silver coating. Since the material of CDCH field wires was Al$5056$ alloy, we started our tests on a sample of Al$5056$ uncoated wires.     
However, the soldering of extremely thin pure aluminium wires is problematic and despite various attempts with different solder pastes and techniques we couldn't obtain enough robust solderings to prevent the wires slipping. Then, we explored the possibility of strengthening the connection of Al$5056$ uncoated wires to the PCBs by using a glue (Loctite $401$). A group of six $\phi=50~\mu{\rm m}$ wires was mounted and glued on a PCB and stretched at equivalent tensions $T \sim\left( 27\textendash30 \right)~{\rm g}$.

\begin{figure}[htb]
	\centering
	\includegraphics[width=1\linewidth]{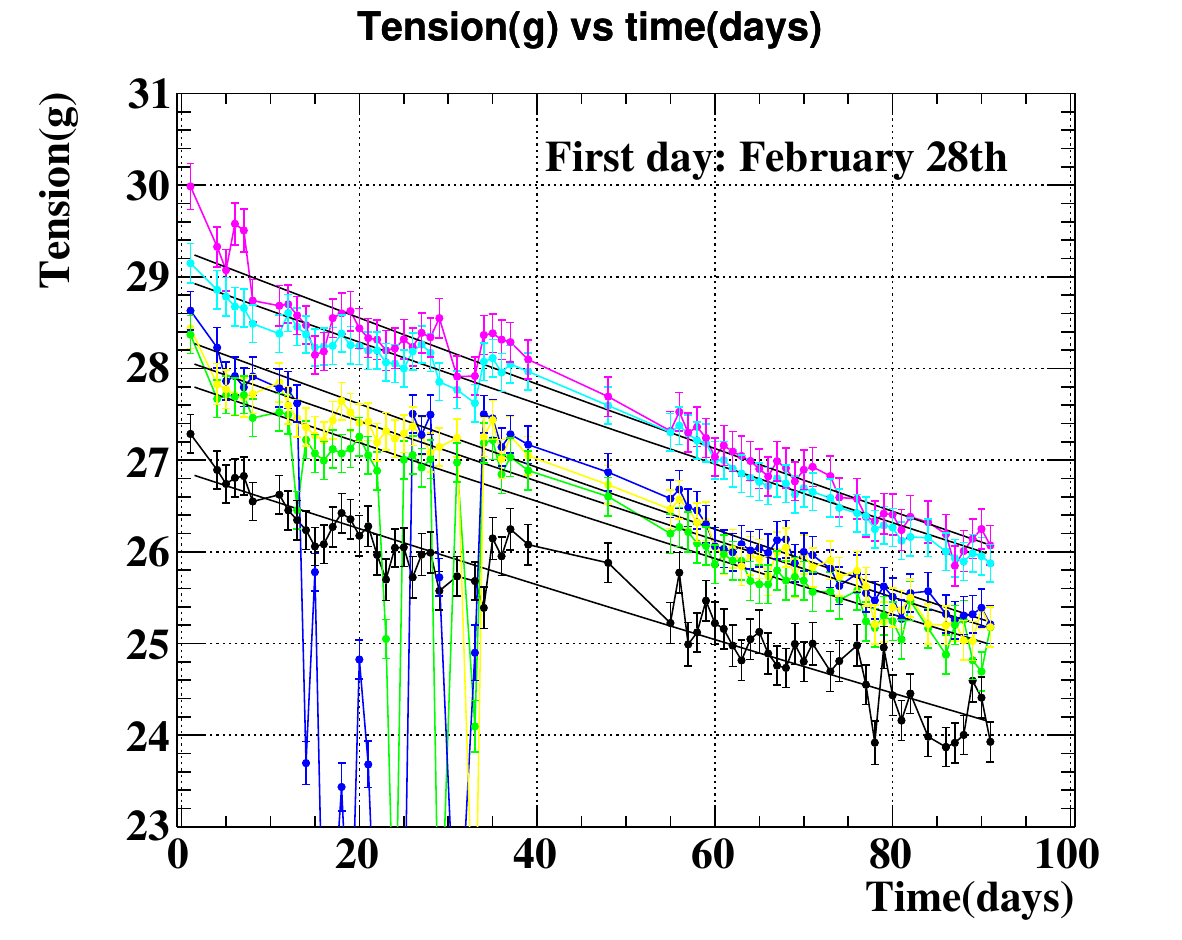}
	\caption{Mechanical tension of six uncoated Al$5056$ wires as a function of time from the glueing date on a PCB; each colour corresponds to a different wire. All curves are fit with the combination of a constant and an exponential decreasing function; the results of the fit are the black lines superimposed to the data. Since all curves are monotonically decreasing, the fit value for the constant component is consistent with zero for all wires.}
	\label{fig:rawfversustimeglued}
\end{figure}

Figure~\ref{fig:rawfversustimeglued} shows the mechanical tension of these wires as a function of time starting from February, $28^{th}$, $2020$, when the wires were connected to the PCB, and ending at the end of May $2020$. 
Each colour corresponds to a different wire and the curves are fit by an exponentially decreasing function and a constant term. The unstable behaviour of the wires corresponding to the blue (and less frequently to the green) dots is due to accidental contacts between the wires and the ground plane which were eliminated after the first weeks of measurements by a more accurate control of the ground plane position. The tension of all wires is continuously decreasing, with a $\sim13\%$ loss in three months, clearly unacceptable. The exponential time constants of the six wires vary between $\sim780$ and $\sim830$ days and all fit values of the constant term are compatible with zero within the fit uncertainties. As we will show later, we observed a dependence of the frequency on the temperature. This effect was measured in $2021$ and in the Spring period whose amplitude was $\le 0.2~{\rm Hz}$. The observed decrease in figure~\ref{fig:rawfversustimeglued} was of $\sim 3~{\rm g}$, corresponding to more than $6~{\rm Hz}$; so, temperature corrections for this set of measurement are negligible. 

The conclusion of this set of measurements is that this glue does not guarantee the required mechanical stability and cannot be the right solution for mounting pure aluminium wires on PCBs. Other types of glue are under testing and one of them (3M-DP 100) is giving promising results; we will discuss this hypothesis in the final section. 

\subsection{Measurements on soldered wires}
The second set of measurements was performed on soldered $\phi=50~\mu{\rm m}$ aluminium wires. Because of the soldering difficulties with the uncoated Al$5056$ wires, we examined unplated wires made by a different alloy, Al$2024$, which has a lower percentage of magnesium and is easier to solder. Indeed we were able to solder a sample of six Al$2024$ wires on a PCB by using a TAMURA ELSOLD Al-S soldering paste and we could test their long term mechanical stability after stretching. The purpose of this set of measurements was to check if, after an initial decrease of the tension due to the creeping, the wire tension reached a saturation value, which remained stable within few tenths of gram, as needed. 
The wires were soldered at the half of June $2020$ and stretched at a tension $\sim24~{\rm g}$; the first data acquisition was performed on June, $15^{th}$, $2020$. The measuring sample was formed by six wires, but one of them is not shown in the following figures because its signal was too low for a reliable measurement of the resonant frequency.

\begin{figure}[htb]
	\centering
	\includegraphics[width=0.495\linewidth]{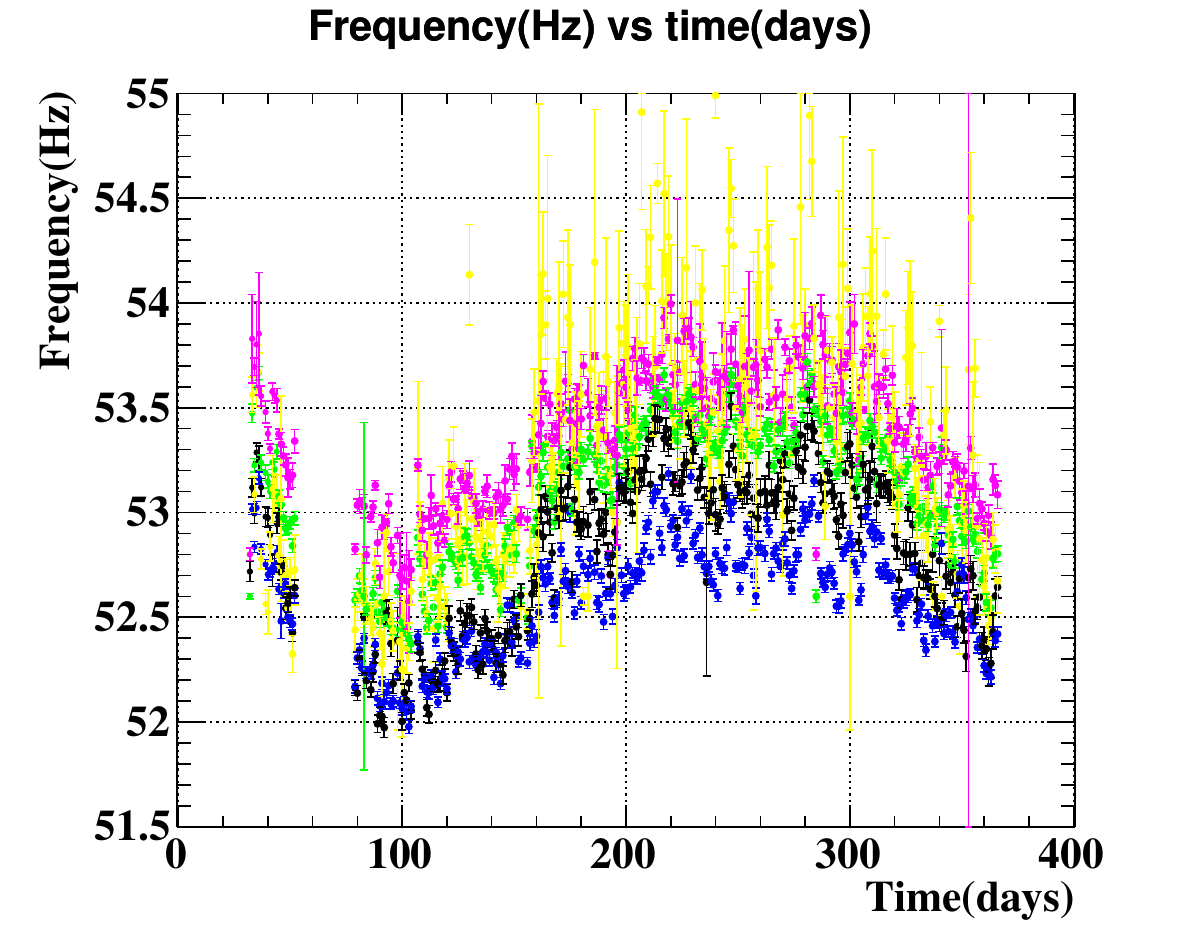}
         \includegraphics[width=0.495\linewidth]{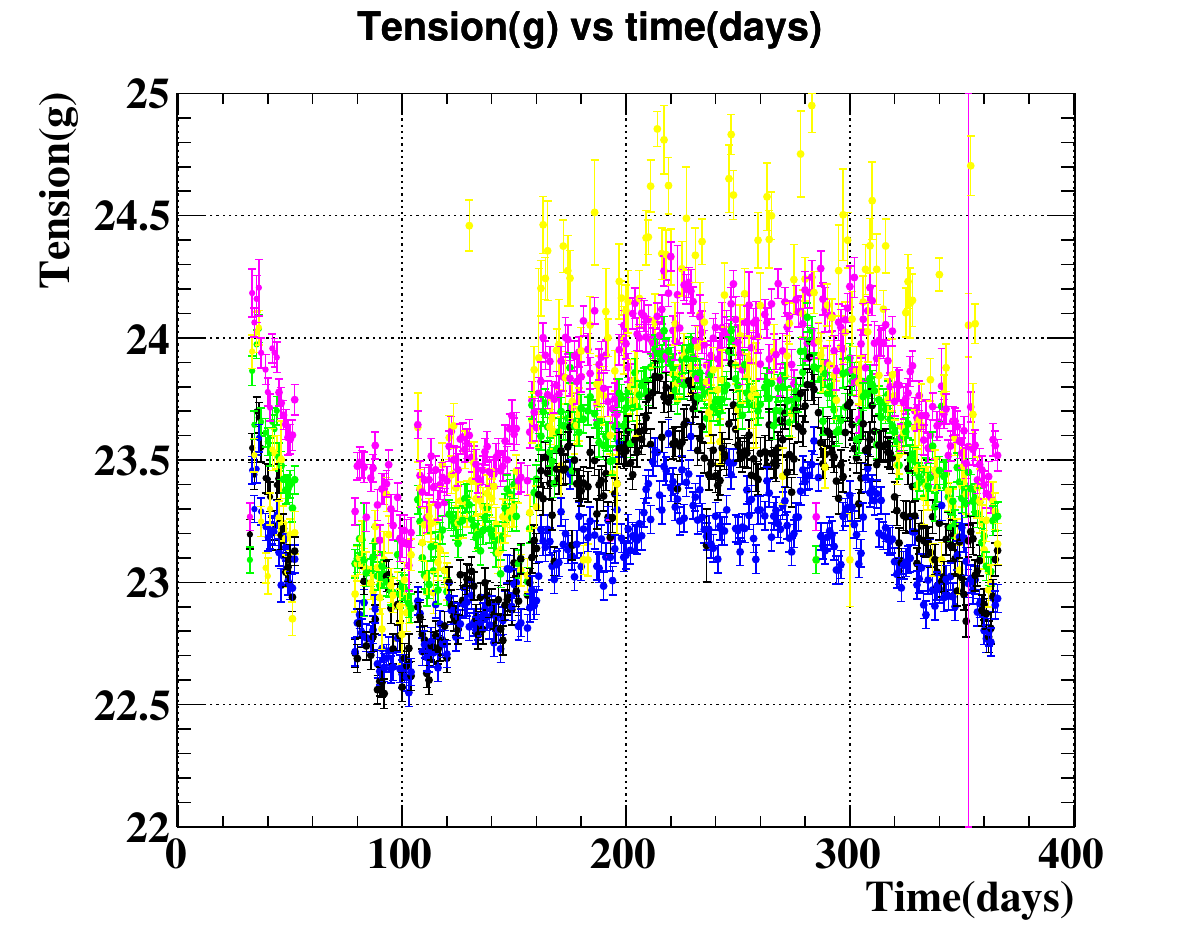}
	\caption{Resonance frequency (left) and mechanical tension (right) of five Al$2024$ wires as a function of time; each colour corresponds to a different wire and the origin of the time axis is June, $15^{th}$, $2020$. Data collected in the first $34$ days are not reliable and are not shown here. Because of its unstable behaviour after day $160$, the \lq\lq yellow wire\rq\rq~was excluded from the global analysis.}
	\label{fig:rawfversustime}
\end{figure}

Figure~\ref{fig:rawfversustime} shows the resonance frequency (on the left) and the mechanical tension (on the right) of the five $\phi=50~\mu{\rm m}$ Al$2024$ wires as a function of time, starting from June, $15^{th}$, $2020$ (day $0$). 
Each colour corresponds to a different wire; the \lq\lq yellow wire\rq\rq~exhibited an unstable behaviour after day $160$ and was excluded by the global analysis which will be discussed later. 
The data of the first $34$ days are not reliable because of an imperfect setting of the ground plane and are not reported here; then, the effective starting time of the data acquisition is day $34$, around the half of July $2020$. There was also a data taking interruption of about $25$ days because of a problem on the electric lines which happened in August $2020$ during one of the most severe phases of the COVID-19 pandemic and was solved at the end of the month. Starting from day $34$, we observed an initial decrease of the resonance frequency (and then of the tension, according to equation \ref{eq:fwt}) with time, with a timing constant $\tau \sim20$ days, followed by an increase starting at day $\sim100$. The first decrease is not unexpected, since wire creeping phenomena are frequently observed when an elastic wire is stretched; however, after a relaxation time of several days one expects that the resonance frequency reaches a stable saturation value; then, the increase is totally unexpected. After day $\sim240$ the resonant frequency and the tension remained at almost stable values, within few tenths of Hz (or g), corresponding to relative fluctuations of $< 1\%$ with respect to the saturation level, for about $3$ months; finally, a decrease of the the resonant frequency with time was observed for day $>330$.   

\begin{figure}[htb]
	\centering
	\includegraphics[width=1\linewidth]{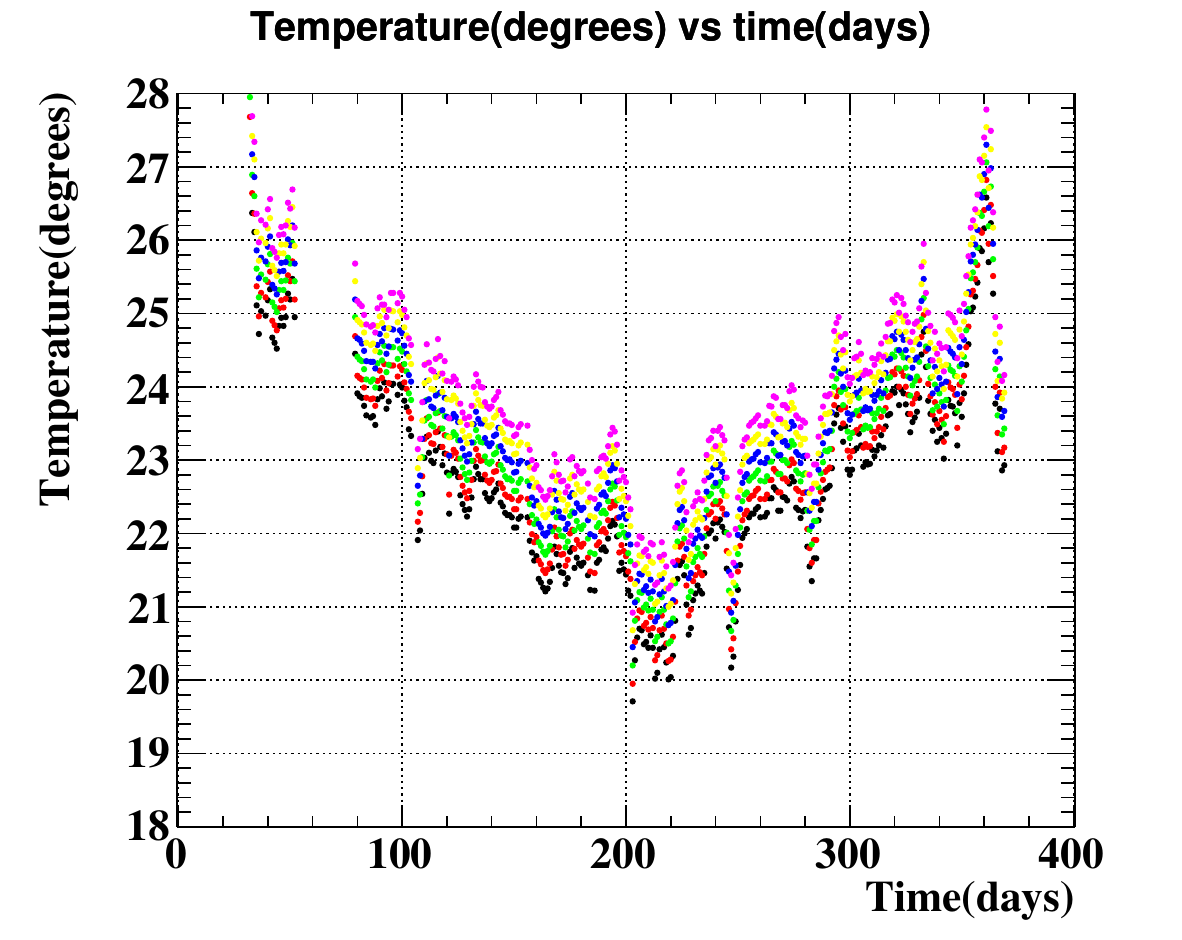}
	\caption{Temperature versus time in the laboratory where the measuring system was located starting from June, $15^{th}$, $2020$. The temperatures recorded during the data acquisition of each individual wire are  shown separately with the same colours used in figure~\ref{fig:rawfversustime} and are shown shifted by $0.25^{\circ} {\rm C}$ on the $y$ axis for clarity.}
	\label{fig:temperature}
\end{figure}

Figure~\ref{fig:temperature} shows the room temperature in the laboratory, recorded during the data acquisition of all wires, as a function of time. The measured temperatures were, day by day, pretty stable during the acquisition time of all the six wires; however, the values recorded during the data acquisition of each individual wire are shown shifted by $0.25^{\circ} {\rm C}$ on the $y$ axis for clarity\footnote{The true temperature values are shown in the black curve. Without the shift, all points corresponding to a single day would be totally superimposed.}.   
As expected, the temperature changed by several degrees according to the different seasons; there was an almost linear decrease with time in Summer, Autumn and Winter $2020$, from day $34$ to day $\sim240$, followed by an increase at the end of Winter and in Spring $2021$.

\begin{figure}[htb]
	\centering
	\includegraphics[width=1\linewidth]{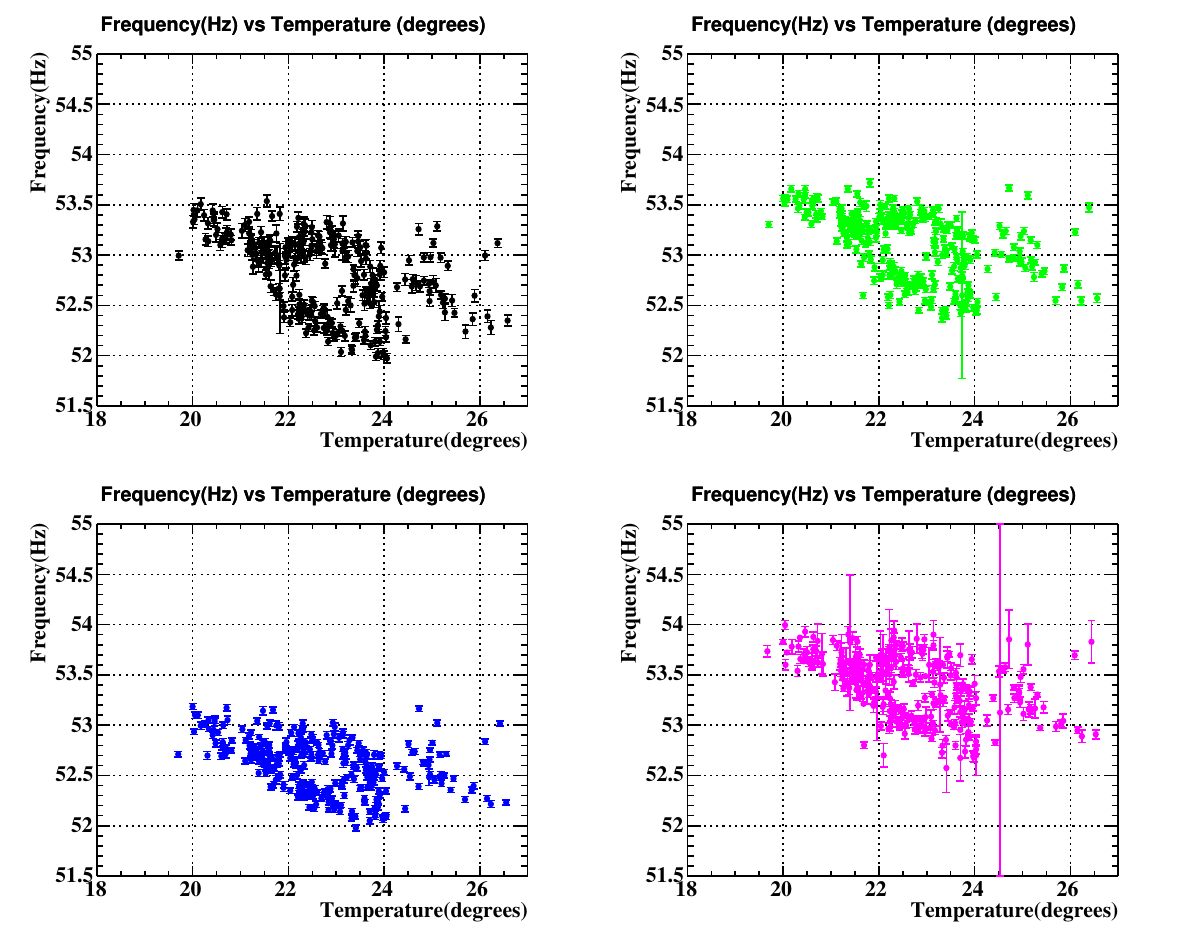}
	\caption{Resonant frequency versus temperature in the laboratory starting from June, $15^{th}$, $2020$ for four wires, identified by the same colour code used in the previous plots.}
	\label{fig:tempfreq}
\end{figure}

\begin{figure}[htb]
	\centering
	\includegraphics[width=1\linewidth]{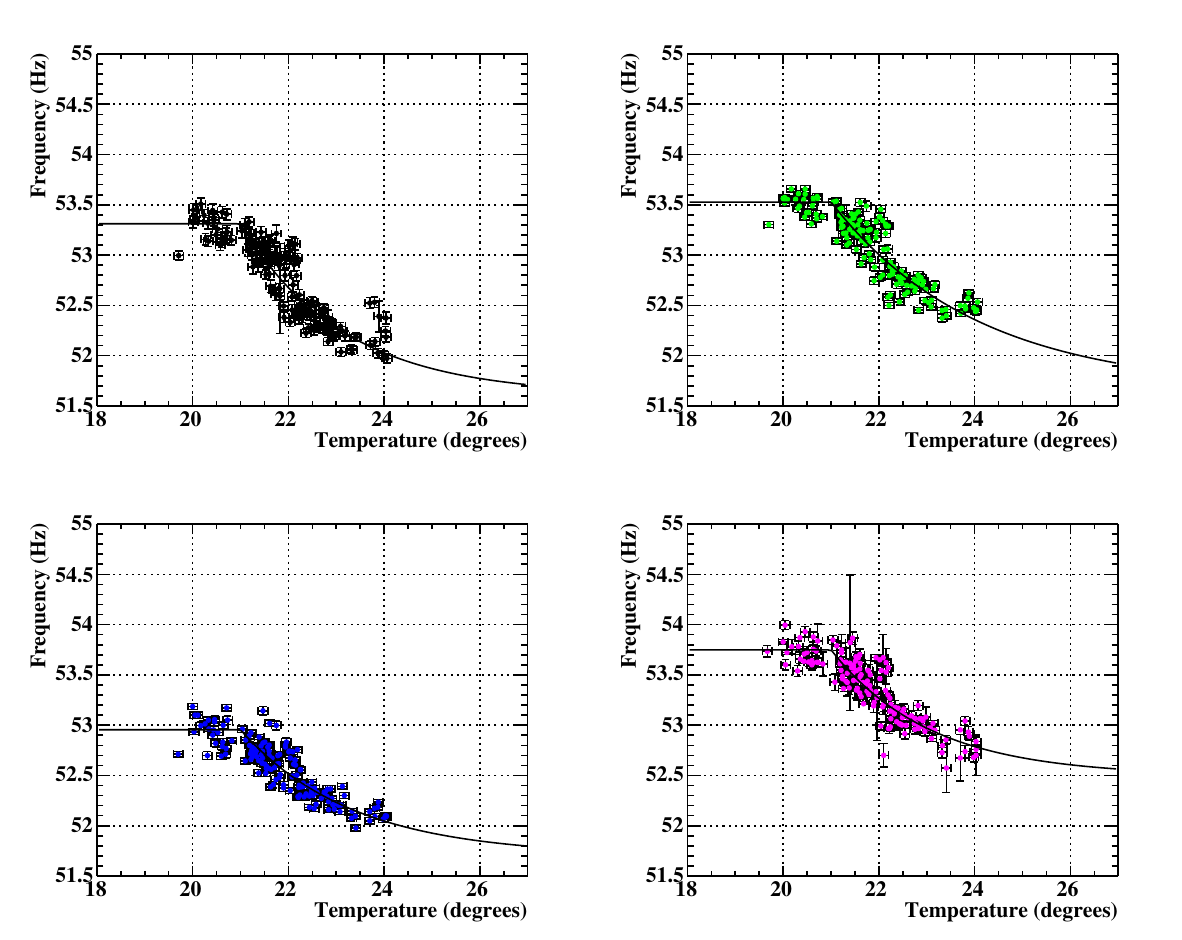}
	\caption{Resonant frequency versus temperature in the laboratory during period b) fit with calibration function \ref{eq:freqversustempcalib}.}
	\label{fig:tempfreqcalib}
\end{figure}

The correlation between the resonant frequency and the temperature for four wires\footnote{The \lq\lq yellow wire\rq\rq~was excluded from this global analysis, as already observed.} is shown in figure~\ref{fig:tempfreq}.
In these plots one can observe four, partially overlapping, patterns:
\begin{itemize} 
\item an isolated group of points for temperatures $> 24^{\circ} {\rm C}$;
\item a flat dependence between $\sim20^{\circ} {\rm C}$ and $\sim21^{\circ} {\rm C}$, followed by a clear decreasing trend in the temperature range $\left( 21\textendash24\right)^{\circ} {\rm C}$;
\item a slowly increasing trend of the frequency for temperatures between 
$\sim21.5^{\circ} {\rm C}$ and $\sim22.1^{\circ} {\rm C}$, followed by a flat dependence 
until $\sim23.2^{\circ} {\rm C}$;
\item an approximately linear drop of the frequency for temperatures between $\sim23.2^{\circ} {\rm C}$ and $\sim24.0^{\circ} {\rm C}$.
\end{itemize} 
These four groups of points correspond to different periods of data taking: 
Summer $2020$, days between $34$ and $55$, before the interruption due to the problem on the  electric lines, from now on: period a); Autumn and Winter $2020\textendash2021$, days between $80$ and $250$, from now on: period b); Spring $2021$, days between $250$ 
and $330$, from now on: period c); late Spring of $2021$, day $> 330$, from now on: period d). According to figure \ref{fig:temperature}, the temperature is almost monotonically decreasing in periods a) and b) and increasing in periods c) and d). The transition between period b) and c) takes place at $T\sim21.5^{\circ} {\rm C}$ for all wires. The relationship between frequency and temperature is, at least qualitatively, similar in the period pairs a)+b) and c)+d): in both cases the frequency is essentially independent of the temperature in a certain range ($< 21^{\circ} {\rm C}$ in the former case, between $21.5^{\circ} {\rm C}$ and $23.2^{\circ} {\rm C}$ in the latter case) and then starts to decrease with increasing temperature; the decreasing slope is faster in period d) than in period b). The origin of this dependence of the frequency with the temperature is probably an electronic problem, as we will discuss in the conclusions. We tried to simulate our electronic chain and we observed such a dependence, but we couldn't reproduce the experimental results quantitatively. Nevertheless, the dependence of the frequency on the temperature must be corrected with some empiric forms in order to obtain a reliable dependence of the frequency (and then of the mechanical tension) as a function of time. This is particularly important in periods a) and b), where one expects to observe an initial decrease followed by a (presumably) flat pattern. Therefore, since the initial decrease of the resonant frequency with time is expected to be complete at the time of period b)\footnote{The starting day of period b) is more than three times larger than the timing constant of the initial decrease.}, we decided to determine a frequency versus temperature correction function by using data collected in this period, where a 
$\sim4^{\circ} {\rm C}$ thermal excursion was observed in almost six months. This correction function will be applied also to period a), in order to disentangle the dependence on the temperature of the frequency vs time curve in the initial phase and single out the wire creeping effects. Figure~\ref{fig:tempfreqcalib} shows the resonant frequency versus temperature curves for the four wires for period b); the curves were fit with the following function:
\begin{equation}
\nu_{r} \left( T \right) = 
	\begin{cases} 	
	\nu_{r,0} + \nu_{r,1} & T < T_{min}  \\
	\nu_{r,0} + \nu_{r,1} \exp{\left[ -\left(\frac{T-T_{min}}{T_{ref}} \right) \right]}  
                                     	    & T \ge T_{min} 
	\end{cases}		
	\label{eq:freqversustempcalib} 
\end{equation}	
where $T_{min}$ is kept fixed at $21^{\circ} {\rm C}$ and $\nu_{r,0}$, $\nu_{r,1}$ and $T_{ref}$ are free fitting parameters. $\nu_{r,0}$ is the resonant frequency when its dependence on temperature becomes negligible and 
$\left( \nu_{r,0} + \nu_{r,1} \right)$ is the resonant frequency at $T \le T_{min}$. All the wires exhibit the same behaviour, with 
$T_{ref} = \left( 0.9\textendash1.2 \right)^{\circ} {\rm C}$ and $\left( \nu_{r,0} + \nu_{r,1} \right) = \left( 52.5\textendash53.5 \right)~{\rm Hz}$. 
The main difference between the wires is the saturation value of the resonant frequency $\left( \nu_{r,0} + \nu_{r,1} \right)$. For what concerns period c), the dependence on the temperature is very weak, despite a temperature range of $\sim 1.7^{\circ} {\rm C}$ in this period, with oscillations of $\sim \left( 0.1\textendash0.2 \right)~{\rm Hz}$ without a clear trend after the initial increasing one, whose maximum amplitude is in any case $\sim 0.2~{\rm Hz}$. Since this effect is well within our tolerance, we decided to avoid further and not well defined corrections and use only the saturation value for period c). Finally for period d) we tried to use a fitting function similar to equation \ref{eq:freqversustempcalib} with $T_{min} = 23.2^{\circ} {\rm C}$, but since the fit was not completely satisfactory we decided to use a simple linear fit whose slope is 
$-\left( 0.3\textendash0.4 \right)~{\rm Hz}/{}^{\circ} {\rm C}$ for $T>23.2^{\circ} {\rm C}$ (each plot has its own slope). Then the correction function $\nu_{r} \left( T \right)$ is given by equation \ref{eq:freqversustempcalib} for periods a) and b), by a constant value for period c) and by the linear fit for period d). After performing the calibration procedure, inserting the corrections and obtaining the complete frequency versus time curve, we will check our assumption that the timing dependence of the resonant frequency in period b) is really negligible.  

\begin{figure}[htb]
	\centering
	\includegraphics[width=1\linewidth]{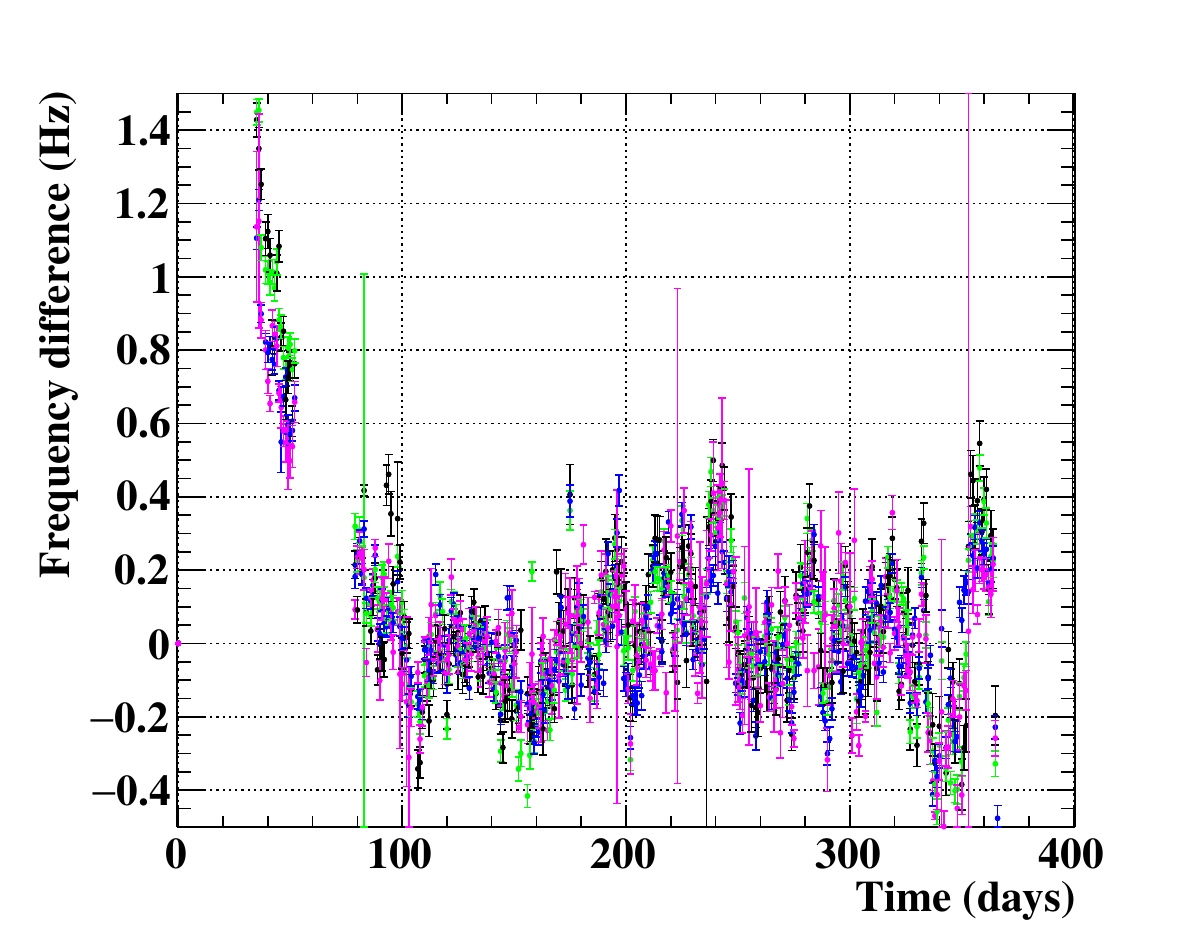}
	\caption{Difference between the measured resonant frequency and the temperature correction function for all the wires superimposed. After subtracting the saturation values the behaviours of all the wires as a function of time are very similar.}
	\label{fig:allfreqversustime}
\end{figure}

In figure \ref{fig:allfreqversustime} we show the difference between the measured frequency and the function $\nu_{r} \left( T \right)$  as a function of time for all wires superimposed.
We used the coefficients deduced from the fits shown in figure \ref{fig:tempfreqcalib} to calculate the temperature dependence of the frequency in periods a) and b) according to equation \ref{eq:freqversustempcalib}, while for period c), even if the average temperature was higher than $T_{min}$, we used the saturation constant $\left( \nu_{r,0} + \nu_{r,1} \right)$ plus a correction $\sim 0.1$~{\rm Hz}, according to the experimental distributions in figure \ref{fig:tempfreq}; finally for period d) we applied the linear correction quoted above and determined by fitting individual plots. 
The effect of introducing the temperature corrections is to recover the expected behaviour of the resonant frequency of all wires with time: an initial, exponential decrease due to the wire creeping, followed by an almost stable resonant frequency. Note that points corresponding to individual wires superimpose very well, confirming that different wires have a different saturation value, i.e. each one has its own resonant frequency, but the same relaxation time. In figure \ref{fig:allfreqversustime} one can observe a possible modulation with time of the resonant frequency. This can suggest that temperature corrections need to be refined or that other effects should be taken into account. We checked if the possible modulation pattern is correlated with the room humidity which is monitored together with the temperature, but no signs of correlations were observed. Nevertheless, the maximum amplitude of resonant frequency oscillations is at level of $0.3~{\rm Hz}$. Table \ref{tab:reson} shows that a $0.3~{\rm Hz}$ uncertainty on the resonant frequency corresponds to a $\left( 0.1\textendash0.2\right)~{\rm g}$ uncertainty on the wire tension, which matches very well our requirement of an uncertainty $< 0.5~{\rm g}$, as explained in section $3$. There are few points with huge error bars, due to sporadic instabilities of one wire which made the resonance signal barely visible above the background and then the quality of the lineshape fit very poor. 

\begin{figure}[htb]
	\centering
	\includegraphics[width=1.0\linewidth]{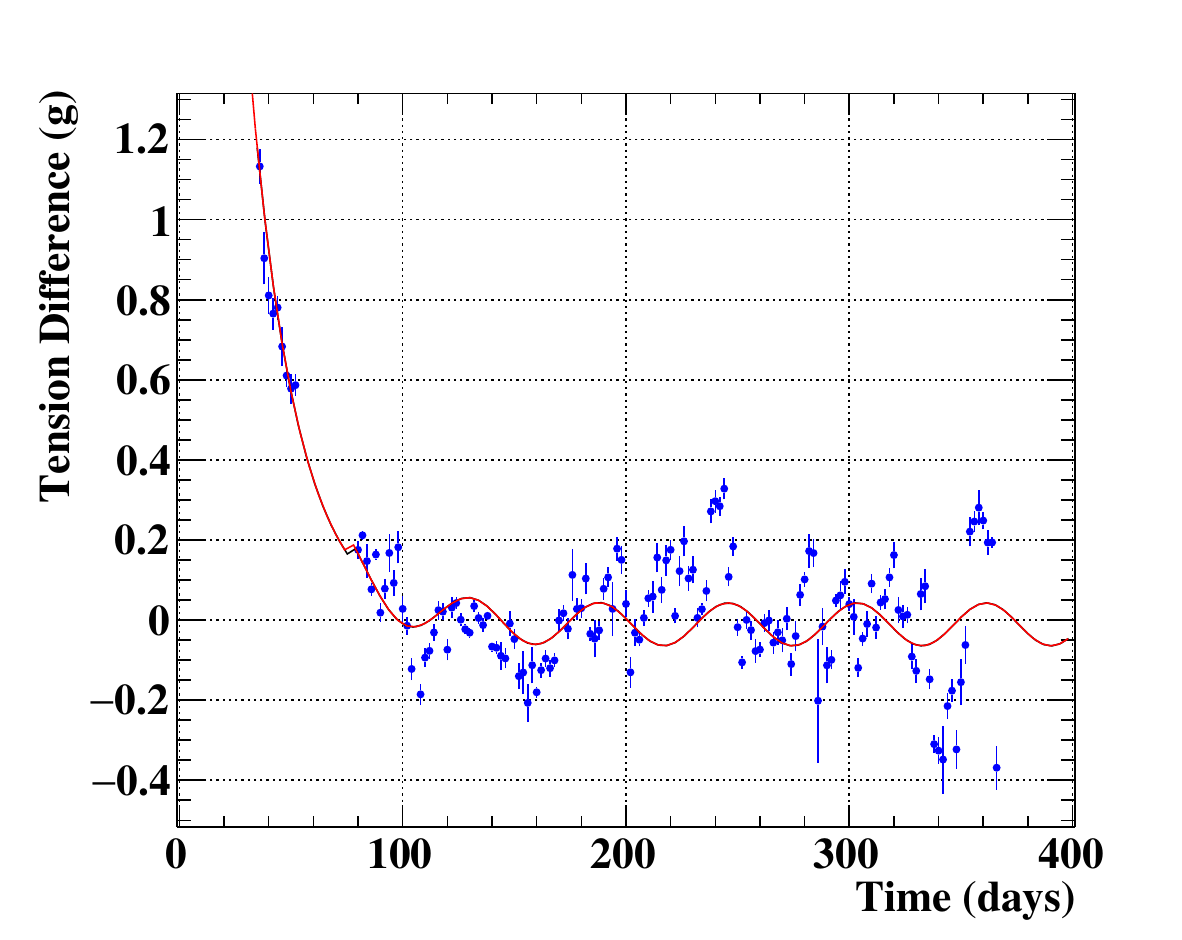}
	%\caption{Difference between the measured resonant frequency (on the left) and the temperature correction function, equation \ref{eq:freqversustempcalib}, summed and averaged over all wires, as a function of time. The plot on the right is the corresponding one for the mechanical tension. The fit function is a combination of a decreasing exponential shape, a saturation constant and an oscillating pattern.}
	\caption{Difference between the measured tension and the temperature correction function, summed and averaged over all wires, as a function of time. The fit function is a combination of a decreasing exponential shape, a saturation constant and an oscillating pattern.}
	\label{fig:fdifferencefit}
\end{figure}

Since the evolution of the resonant frequency with time is very similar for all wires, we converted the values from frequency to tension, repeated for tension points the calculations used in figure \ref{fig:allfreqversustime} for frequency points, summed and averaged the experimental points and fitted the average tension versus time relationship with the combination of an exponential shape, a saturation constant and an oscillating pattern. The results are shown in figure \ref{fig:fdifferencefit}.
The fit parameters (exponential time constant, saturation constant, amplitude and period of the oscillating pattern) are reported in table \ref{tab:finalfit}.
\begin{table}[htb]
	\caption{Parameters of the fit function shown in figure \ref{fig:fdifferencefit}.} 	
	\centering
	\begin{tabular}{|c|c|c|}
		\hline
		Parameter & Parameter value & Units \\
		\hline
		\hline
		%Exponential time constant & $21.0 \pm 0.5$                                       & days \\
		%Saturation constant          & $\left( -25 \pm 30 \right) \times 10^{-3}$ & Hz    \\
		%Modulation amplitude       & $\left( 57 \pm 3 \right) \times 10^{-3}$    & Hz    \\
		%Modulation period            & $51 \pm 5$                                             & days \\
		Exponential time constant & $21.1 \pm 0.5$                                       & days \\
		Saturation constant          & $\left( -11 \pm 3 \right) \times 10^{-3}$   & g    \\
		Modulation amplitude       & $\left( 46 \pm 3 \right) \times 10^{-3}$    & g    \\
		Modulation period            & $51 \pm 5$                                             & days \\
		\hline
	\end{tabular}
	\label{tab:finalfit}
\end{table}
At the beginning of period b) the contribution of the exponential term is $\sim0.07~{\rm g}$, confirming our assumption that the timing dependence of the resonant frequency (and therefore of the tension) in the timing interval used to determine the temperature correction was marginal with respect to the temperature dependence. The saturation constant is $\sim-0.01~{\rm g}$, showing that the temperature corrections are adequate, at least on average. The modulation amplitude is $46~{\rm mg}$ in the frequency, largely within our precision requirements on the knowledge of the wire tension. The origin of the modulation is not yet explained but, as already noted, it has no significant effects on the accuracy of the measurement. We tried also a simpler fit without the modulation term; the results for the exponential time constant and the saturation constant are in good agreement with that reported in table \ref{tab:finalfit}. Including the modulation term in the fit has the advantage of providing a quantitative estimate of the wire tension stabilities; both look pretty stable within small fluctuations on a  timing scale of (at least) several months.  

\begin{figure}[htb]
	\centering
	\includegraphics[width=1\linewidth]{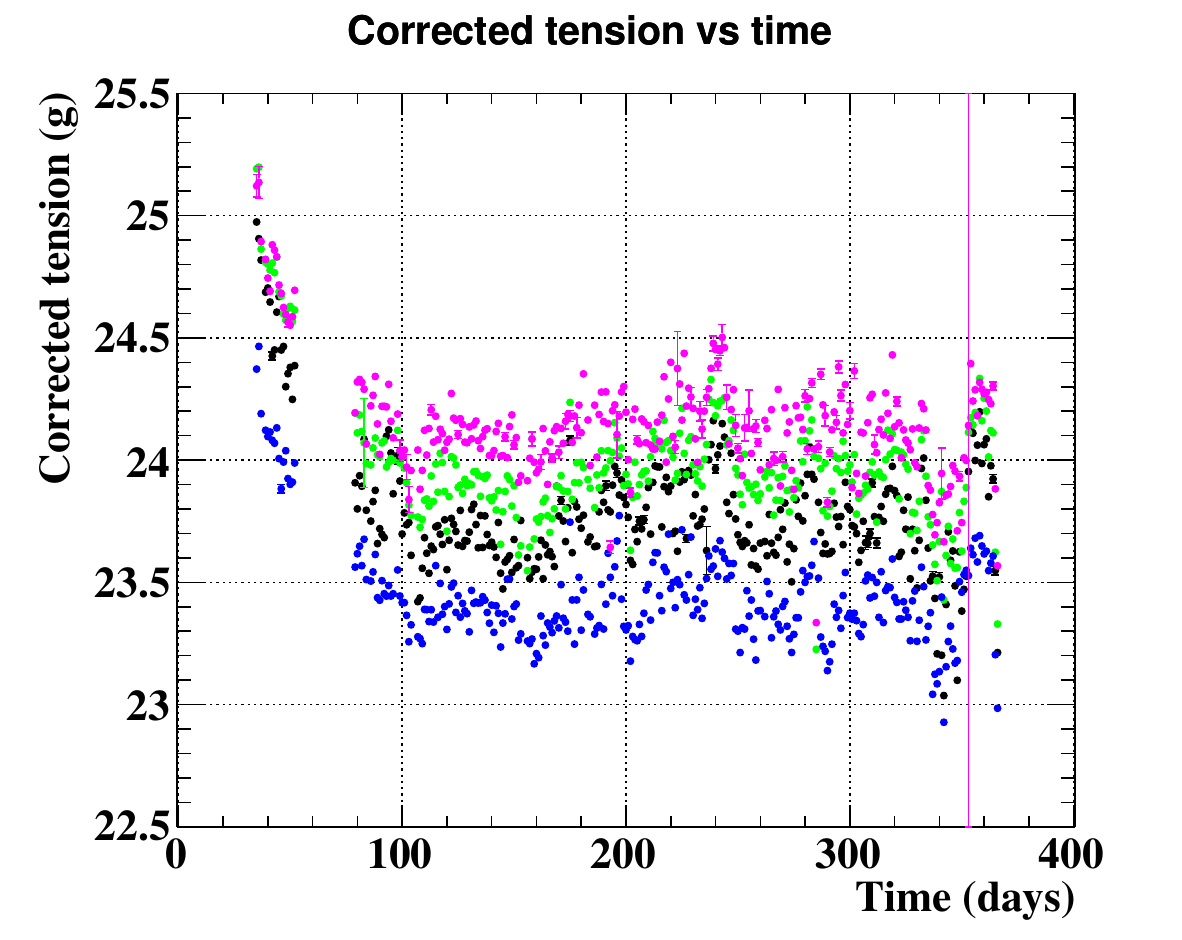}
	\caption{Mechanical tension of four soldered Al$5056$ wires as a function of time from the soldering date on a PCB, after the application of the temperature corrections; each colour corresponds to a different wire. After the initial decrease due to the creeping, the tensions of all wires remain stable within $< 0.2~{\rm g}$.}
	\label{fig:tensvstimesold}
\end{figure}

Finally we show in figure \ref{fig:tensvstimesold} the mechanical tension of the four Al$2024$ soldered wires with time after the application of temperature corrections. 
As one can see, after the initial (expected) decrease, all wires exhibit a stable behaviour of the mechanical tension as a function of time, with oscillation amplitudes $< 0.2~{\rm g}$, which fit well our required precision. Moreover, the distribution of the tension values for day $> 100$ is, wire by wire, pretty gaussian, with a standard deviation in the range $\left( 0.12\textendash0.15 \right)~{\rm g}$, again well within our requests. 
\subsection{Comments on the choice of the wire material for the new CDCH}
On the basis of the measurements shown in the previous subsection the Al$2024$ uncoated wires appear to have good mechanical properties to be used for the new CDCH. However, this kind of wires didn't pass a corrosion test, since some Al$2024$ wires were immersed in distilled water and broke in many small pieces in less than two days, without need of stretching. The alloy Al$2024$ was then rejected as a possible building material for the wires of the new CDCH and we came back to the silver plated Al$5056$ wires, whose mechanical properties were already tested during the assembly phase of the old CDCH. Since the breaking probability of the $\phi = 50~\mu{\rm m}$ wires was much smaller than that of $\phi = 40~\mu{\rm m}$ wires, we deciced to use only the more robust thicker diameter wires and after discussions with the manifacturers they agreed to avoid the ultrafinishing stage in the wire production and started the fabrication of non-ultrafinished spools which are under testing.

\section{Conclusions}
The measurement system of the resonant oscillation frequency of thin metallic wires developed for the MEG II experiment measures a small variation (a few pF) of the capacitance of a wire with respect to a ground plane induced by a sinusoidal HV signal.
An oscillation amplitude versus frequency curve is obtained by means of a frequency scan and is fitted by a Breit-Wigner shape to extract the resonance frequency; the measured wire frequency can be converted into the wire tension and the corresponding elongation. 

After the wire breakage problems due to chemical corrosion observed in the MEG II CDCH,  the construction of a new chamber was approved. The original chamber was in any case completed after removing all fragments of broken wires and it is taking data at PSI; nevertheless, the project of the new chamber went on and the fundamental point to be addressed before the start of its constructon is the type of wire to be chosen for replacing the too fragile $\phi=40~\mu{\rm m}$ silver plated aluminium wires. Several possibilities were explored and the measuring system described in this paper was a very useful tool to check the reliability of some of the possible solutions. 

First of all, the possibility of using uncoated Al$5056$ aluminium wires was tested, but it 
appeared problematic because of the difficulty of soldering these wires on PCBs and because the backup solution of ensuring the mechanical connection between the wires and the PCBs by means of a glue was not satisfactory: we observed a continuous loosening of the wire tension, corresponding to a $13\%$ loss in less than three months, without any indication of reaching a stable value.    

Secondly, we examined the possibility of a different aluminium alloy, Al$2024$, easier to solder even without silver coating. After soldering a group of Al$2024$ $\phi=50~\mu{\rm m}$ wires on a PCB, we studied the timing behaviour of their resonant frequency and mechanical tension on a time scale of about one year. We observed an initial tension loosening due to the creeping, as usual when a thin mechanical string is stretched, followed by a long term stability, with some (probably artificial) residual oscillations, whose amplitude is well within our tolerance. However, the Al$2024$ wires were rejected because of their weak resistance to the humidity induced corrosion and we reverted to $\phi=50~\mu{\rm m}$ silver plated wires made with the original Al$5056$ alloy, but avoiding the ultrafinishing 
stage which produces a strong mechanical stress on the silver coating.  

Despite the impossibility of using Al$2024$ wires, the long time measurements on this kind of wires and on the glued Al$5056$ wires showed the potentialities and the performances of our measuring system. The resonant frequency of the wires can be measured with a precision of few tenths of Hz, corresponding to a precision on the knowledge of the mechanical tension $\sim\left( 0.1\textendash0.2 \right)~{\rm g}$, well within our requirements. The measuring system operates with very small human interventions, can be controlled remotely (an important \lq\lq quality factor\rq\rq~during the COVID-19 pandemic) and a complete set of measurements requires $\sim\left( 15\textendash20\right)$ minutes per day of data acquisition; also the analysis chain is largely automated, very fast and based on standard packages. The recording of environment conditions, room temperature and humidity, simultaneuously to the data acquisition, allows to correct for the dependence of the resonant frequency on the room temperature, a crucial step to obtain reliable frequency versus time curves. The origin of the dependence of the measured resonant frequency on the room temperature is probably the inverter chain of the Ring oscillator, which is highly sensitive to the temperature. A redesign of the Ring oscillator circuit would be necessary to correct this dependence at hardware level; presently this solution is under discussion. Another simpler strategy for improving the measurement quality is to move the experimental apparatus in a temperature and humidity controlled room. 

The final decision of the wire type for the new chamber is not yet taken. The main option is to use non ultrafinished silver coated Al$5056$ wires, but we found recently an efficient way of soldering pure aluminium wires of Al$5056$ type with the TAMURA ELSOLD Al-S soldering paste which is able to sustain a tension up to $60~{\rm g}$ without wire slipping. In this case it is safer to reinforce the wire  mechanical connection by means of a glue; we are testing the mechanical strength and the chemical properties of the 3M DP100 glue, which seems a good candidate for our purposes.  

The measuring system discussed in this paper will be used for long time stability tests of these and other possible choices of wire types and connections with PCBs. 
\section{Acknowledgements}
{We are very grateful to several technicians of our universities and INFN Sections: G.Balestri, A.Bianucci, M.D'Elia, A.Innocente, A.Miccoli, C.Pinto, G.Petragnani and A.Tazzioli who were deeply involved in the CDCH wiring and mounting procedures.}

\end{document}